\documentclass[%
reprint,
superscriptaddress,
 amsmath,amssymb,
 aps,
prl,
]{revtex4-1}

\usepackage{graphicx}
\graphicspath{ {Graphs/} }
\usepackage[usenames,dvipsnames,svgnames,table]{xcolor}
\usepackage{dcolumn}
\usepackage{bm}
\usepackage{tabularx}

\usepackage{hyperref}

\usepackage{epstopdf}
\usepackage{float}
\usepackage{booktabs} 
\usepackage{array} 
\usepackage{paralist} 
\usepackage{verbatim}
\newcommand{\chem}[1]{\ensuremath{\mathrm{#1}}}

\newcommand{\units}[1]{\ensuremath{\,\mathrm{#1}}}
\newcommand{\NiBi}{\chem{NiBi_3}}
\newcommand{\Tc}{\ensuremath{T_c}}
\newcommand{\degC}{\ensuremath{^\circ\units{C}}}
\usepackage{fancyhdr} 
\begin{document}

\title{On the Origin of Superconductivity at Nickel-Bismuth Interfaces}

\author{Matthew~Vaughan}
\affiliation{School of Physics and Astronomy, University of Leeds, Leeds, LS2 9JT, United Kingdom}

\author{Nathan~Satchell}
\affiliation{School of Physics and Astronomy, University of Leeds, Leeds, LS2 9JT, United Kingdom}

\author{Mannan~Ali}
\affiliation{School of Physics and Astronomy, University of Leeds, Leeds, LS2 9JT, United Kingdom}

\author{Christian~J.~Kinane}
\affiliation{ISIS Neutron and Muon Source, Rutherford Appleton Laboratory, OX11 0QX, United Kingdom}

\author{Gavin~B.~G.~Stenning}
\affiliation{ISIS Neutron and Muon Source, Rutherford Appleton Laboratory, OX11 0QX, United Kingdom}

\author{Sean~Langridge}
\affiliation{ISIS Neutron and Muon Source, Rutherford Appleton Laboratory, OX11 0QX, United Kingdom}

\author{Gavin~Burnell}
\email{g.burnell@leeds.ac.uk}
\affiliation{School of Physics and Astronomy, University of Leeds, Leeds, LS2 9JT, United Kingdom}

\date{\today}

\begin{abstract}

Unconventional superconductivity has been suggested to be present at the interface between bismuth and nickel in thin-film bilayers. In this work, we study the structural, magnetic and superconducting properties of sputter deposited Bi/Ni bilayers. As-grown, our films do not display a superconducting transition, however, when stored at room temperature, after about 14\units{days} our bilayers  develop a superconducting transition up to 3.8\units{K}. To systematically study the effect of low temperature  annealing on our bilayers, we perform structural characterization with X-ray diffraction and polarized neutron reflectometry, along with magnetometry and low temperature electrical transport measurements on samples annealed at $70\degC$. We show that the onset of superconductivity in our samples is coincident with the formation of ordered \NiBi\ intermetallic alloy, a known $s$-wave superconductor. We calculate that the annealing process has an activation energy of $(0.86\pm 0.06)\units{eV}$. As a consequence, gentle heating of the bilayers will cause formation of the superconducting \NiBi\ at the Ni/Bi interface, which poses a challenge to studying any distinct properties of Bi/Ni bilayers without degrading that interface.


\end{abstract}


\maketitle

\section{Introduction}

Superconductivity and ferromagnetism are normally considered incompatible phases as the strong exchange field of a ferromagnet will act to break superconducting Cooper pairs \cite{Ginzburg1956}. It is therefore unusual to find a superconducting transition at about 4~K in Bi/Ni bilayers, when ferromagnetic Ni has no known such transition, and crystalline Bi is only superconducting below 0.5~\units{mK}  \cite{Prakash17}. Higher critical temperatures in Bi have been reported under certain conditions, for example,  \Tc\ $\approx 6\units{K}$ in amorphous Bi, \Tc\ $\approx4\units{K}$ induced under pressures of a few \units{GPa}, a \Tc\ range of between 2-5.5\units{K}\ on the surface of grain boundaries and \Tc\ of 1.3\units{K} in nanowires \cite{petrosyan1974,Buckel65,Weitzel91,Li17,Tian09}. None of these can, however, explain the superconductivity in Bi/Ni bilayers.

Bi/Ni bilayer superconductivity was initially discovered in Bi layers grown on a dusting of Ni in tunneling measurements, which showed that the superconductivity extends across the entire thickness of the Bi \cite{Moodera1990}. Later, similar measurements showed that in such bilayers, superconductivity and ferromagnetism coexist \cite{LeClair2005}. Recently, there is a renewed interest in Bi/Ni bilayers as the combination of superconductivity, ferromagnetism and strong spin-orbit coupling may lead to exotic new physics. In particular, epitaxial bilayers of Bi/Ni grown by molecular beam epitaxy have been heavily studied \cite{Gong15,Gong16,Wang17,Zhou16,Zhou17,Chauhan19,tokuda2019spin}. There is speculation that results on these epitaxial bilayers show $p$-wave superconductivity \cite{Gong15}, time-reversal symmetry breaking \cite{Gong16}, and chiral superconductivity \cite{Wang17}.

An alternative explanation for the origin of the superconductivity in the Bi/Ni bilayer is by the presence of the alloy \NiBi\, which is established to superconduct with a similar \Tc\  of 4~K \cite{Alekseevskii1951, Sive15}. Measurements on bulk crystals of \NiBi\  suggest that it shows coexistence of superconductivity and ferromagnetism \cite{Silva13}, however it is expected to be a singlet, $s$-wave, superconductor \cite{zhao18}. Silva \textit{et al} observed this alloy in thin-film Bi-Ni interfaces, which they attribute to spontaneous formation during sample growth at a temperature of $60\degC$ \cite{Silva13}. Liu\textit{et al} also observe interdiffusion during sample growth at $300\units{K}$ but no interdiffusion when samples are grown colder than $110\units{K}$ \cite{Liu2018}. Whilst formation during growth will be dominated by comparatively rapid surface diffusion, it is also important to establish whether, and under what conditions, formation of \NiBi\ can occur post-growth when the as-grown samples show initially clean and distinct interfaces of Bi/Ni.

In this work, we set out to determine at what timescales and temperatures intermixing from an initially distinct Bi/Ni interface becomes significant. Annealing at low temperatures, we use SQUID magnetometry, x-ray diffraction (XRD), and polarized neutron reflectometry (PNR) to measure changes in Bi/Ni samples. X-ray reflectometry is not used as the x-ray scattering density for bulk Bi and NI are too similar for effective contrast at the interface. We observe the onset of a superconducting \Tc , evolution of the magnetic moment, and changing structure of the interface. Our results suggest that special handling of the Bi/Ni samples and refrigerated storage is necessary to prevent the formation of \NiBi , which otherwise occurs after a few days at room temperature, or a few minutes at temperatures typical of many cleanroom processing steps.

\section{Methods}

Samples are grown by DC sputtering from pure metal targets of Bi (99.99\%) and Ni (99.95\%) at ambient temperature ($21\degC$) on to thermally oxidized Si substrates. To prevent oxidation, the samples are capped with 5\units{nm} thick Ta layers. It is found that optimal layer thicknesses for maximum \Tc\ are 50\units{nm} of Bi and 6\units{nm} of Ni, consistent with previous work \cite{Gong15}. See the Supplemental Material for full details of the deposition and \Tc\ dependence on Bi and Ni thicknesses \cite{NoteSM}. 

To prevent unintentional annealing, directly after removing from the deposition system, the samples are stored in a freezer at $\approx -20\degC$ and transported in a portable refrigerator at $\approx 4\degC$. We either anneal our samples at room temperature (21\degC ), or perform controlled annealing of our samples between 50\degC\ and 150\degC\ on a hotplate under a cover to maintain a uniform temperature. 

Polarized neutron reflectometry (PNR) is performed on the PolRef beamline at the ISIS neutron and muon source. Polarized neutron reflectometry data are analyzed using the GenX software \cite{bjorck_genx:_2007}. X-ray diffraction (XRD) is performed on a Rigaku SmartLab diffractometer using Cu K$_{\alpha}\ \lambda=1.54$\AA\ radiation and magnetization loops are measured using a Quantum Design MPMS 3 SQUID magnetometer both courtesy of the ISIS R53 characterization lab. We also employ an Oxford Instruments MagLab 8T VSM for additional magnetization measurements for the temperature dependent annealing. Electrical transport measurements are performed using a standard 4-point probe AC method utilizing a lock-in amplifier and a $77\units{Hz}, 100\units{\mu A}$ current inside a 4He variable temperature cryostat with a 3\units{T} superconducting magnet in a horizontal Helmholtz Coil configuration.

\section{Results}

\subsection{Polarized Neutron Reflectometry Measurements}
\begin{figure*}
  
  \includegraphics[width=0.9\linewidth]{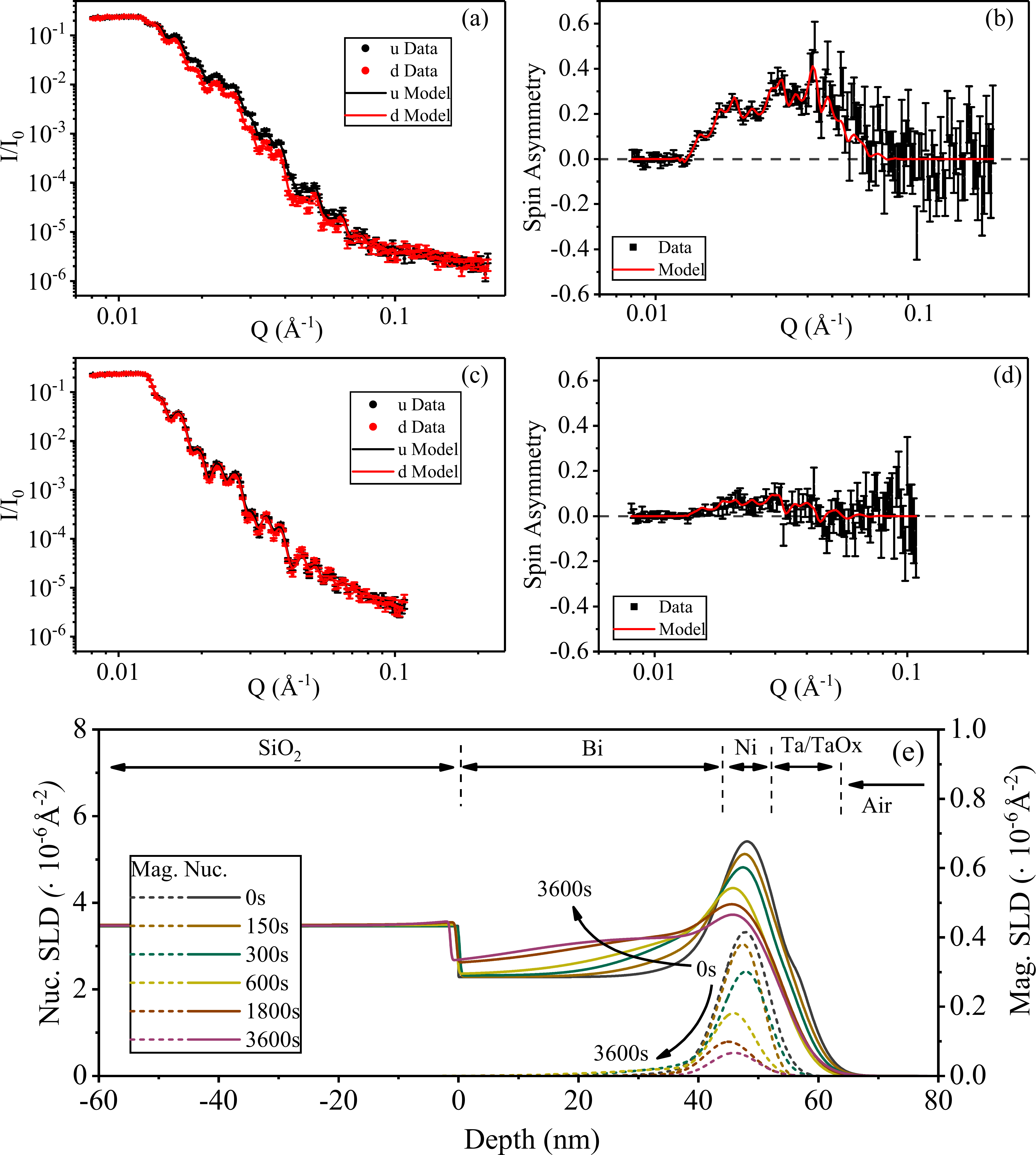}%
  \caption{Polarized neutron reflectometry (PNR) measurements at 10~K and 0.2~T of Si/SiO (100 nm)/Bi (50 nm)/Ni (6 nm)/Ta (5 nm) sample.  (a),(b) PNR and spin asymmetry in the as-grown state. (c),(d) PNR and spin asymmetry after 3600\units{s} annealing at 70\degC. The solid lines are fits to the data and the returned fit parameters are given in Table \ref{Fig:GenX}. (e) The nuclear and magnetic scattering length density (SLD) with depth returned from fitting showing the evolution of SLD with successive annealing.}
  \label{Fig:PNR}
\end{figure*}

By measuring the neutron reflectivity as a function of the wavevector transfer and neutron spin eigenstate, PNR allows the scattering length density (SLD) to be obtained. Careful fitting to the two obtained reflectivity curves in PNR allows the extraction of depth dependent magnetization and structure. PNR is widely employed in the successful characterization of spintronic materials \cite{Zabel2008} and is particularly useful in this study due to the large SLD contrast between the Ni and Bi layers. The sample is annealed repeatedly at 70\degC. Films are loaded at room temperature and cooled in an 0.2\units{T} field to 10\units{K} where we perform PNR measurements.

Figure \ref{Fig:PNR} (a,c) shows the obtained PNR curves for the sample with the corresponding fit to each spin state. (a) shows the as-grown state, while (c) is after 3600\units{s} annealing at 70\degC. The higher frequency oscillations arise from the thick SiO$_{2}$ layer on the substrate. The thinner, rougher, layers of the film modulate these oscillations. The spin asymmetry as calculated by $\frac{u-d}{u+d} $ is shown in Figure \ref{Fig:PNR} (b,d). The spin asymmetry directly scales with the sample's magnetization. The spin asymmetry shows significant reduction between the as-grown and annealed states indicating a reduction of magnetization.

We employ a box model to fit the PNR data with the layers Ta$_{2}$O$_{5}$/Ta/Ni/\NiBi\ /Bi/SiO$_{2}$/Si. Ta$_{2}$O$_{5}$ is the most common oxide of Ta and the SiO$_{2}$ density is taken as amorphous. Each layer is assigned a thickness, roughness, magnetization, scattering length, and density. The scattering lengths are fixed from the known bulk values. Magnetism in the structure is limited to the Ni and \NiBi\ layers, \NiBi\ is not ferromagnetic and we do not find a significant moment inside this layer. All other parameters except the substrate density and thickness (the substrate is infinitely thick) are free fitting parameters with physically realistic bounds defined from either the known bulk or as-grown values. The results of fitting are shown in Figure \ref{Fig:PNR} (e), Table (Fig.\ref{Fig:GenX}) and the full fitted parameters are listed in the Supplemental Material \cite{NoteSM}.

The results of fitting are shown in Figure \ref{Fig:PNR}.(e). As grown the samples do show a thin (2\units{nm}) \NiBi\ layer that has either formed during growth, or in the time taken to remove them from the deposition system. Consequently, the Ni and Bi layers are slightly thinner than expected (4.4\units{nm} and 44\units{nm} versus the nominal 6\units{nm} and 50\units{nm}). The thickness of the \NiBi\ layer in the model is less than the roughness of any of the Bi, \NiBi\ or Ni interfaces suggesting that it is an intermixed region rather than a distinct layer. The Magnetic SLD of bulk Ni is $\approx 1\times 10^{-6}\units{\AA^{-2}}$ where our magnetic SLD for the Ni in the as-grown state is $\approx 0.4\times 10^{-6}\units{\AA^{-2}}$ as the roughness distributes the Ni over a wider range. The area under the Magnetic SLD and the area derived from the box model with no roughness is $3.98\times 10^{-6}\units{\AA^{-1}}$ and $3.97\times 10^{-6}\units{\AA^{-1}}$ respectively.

The main changes to the structure with annealing the sample are the diffusion of Ni atoms into the Bi layer. The stronger nuclear scattering of the Ni increases the SLD of the Bi layer where atoms have diffused, Figure \ref{Fig:PNR}.(e). The structural profile of the final annealed state suggests that the Ni can diffuse across the entire thickness of the Bi as the SLD of the entire layer is raised. At the Bi/Ni interface, we report that a layer with the correct SLD for ordered \NiBi\ intermetallic layer is found. The thickness of this layer increases with annealing from (2\units{nm}) in the unannealed state, to (38.8\units{nm}) in the final annealed state.

As the Ni diffuses; less ferromagnetically ordered Ni is left, in turn decreasing the magnetic SLD profile and moment measured independently from SQUID. For the final two annealing steps, the fitting returns a very thin layer of Ni (0.7\units{nm}). The large roughness (4\units{nm}) of this layer suggests it is no longer continuous and has been replaced by an intermixing layer. In an attempt to improve the modeling, we try replacing this Ni layer with a layer approximating a NiBi intermixing layer. Doing so, we find the fitting returns a lower figure of merit, indicating that model has a closer resemblance to the physical sample. This suggests that in the final annealed state, Bi contaminates the Ni layer in addition to the Ni diffusion into the Bi layer. There is very little change to the scattering length density shown Figure \ref{Fig:PNR} (e) between the two models as the roughness tends to smear such fine details. The full fitting parameters from each model are provided in the Supplemental Material \cite{NoteSM}. 

\begin{table}
  \begin{tabular}{c|c|c}
  Layer (nm) & Initial & Final \\
  \hline\hline
  Ta$_{2}$O$_{5}$ & $1.6\pm 0.3$ & $4\pm 0.5$ \\
  Ta & $4.9\pm 0.4$ & $0.5\pm 0.5$ \\
  Ni & $4.5\pm 0.7$ & $0.7\pm 0.1$ \\
  \NiBi\ & $2.3\pm 0.4$ & $38.8\pm 0.7$ \\
  Bi & $44.1\pm 0.4$ & $9.3\pm 0.7$ \\
  SiO$_{2}$ & $95.0\pm 0.3$ & $93.9\pm 0.5$ \\
  \end{tabular}

  \caption{The initial (0\units{s}) and final (3600\units{s}) annealed fitted thicknesses at 70\degC . The full fitting parameters are given in the Supplemental Material \cite{NoteSM}.}
  \label{Fig:GenX}
\end{table}

\subsection{Magnetometry and X-Ray Diffraction}


To study the magnetic and structural properties of our Bi/Ni bilayer samples we employ SQUID magnetometry and Cu K$_{\alpha}\ \lambda=1.54$\AA\ x-ray diffraction (XRD). We use a sample grown in the same vacuum cycle as the sample we study by PNR, which we dice into smaller (4x4\units{mm}) cuttings. The individual cuttings are treated with the same annealing process at 70\degC~as the PNR sample. After annealing, we first measure the magnetic hysteresis (moment vs field) of the cuttings at 10~K (the same temperature as the PNR), then check for superconducting transition by measuring moment vs temperature from the base temperature of the magnetometer at small applied field (2-10~K sweep at 5~mT), finally we transfer the cuttings to the x-ray diffractometer where XRD is measured at room temperature. To minimize annealing during the x-ray measurements, the total time to align and record a XRD scan is optimized to take about 30 minutes. Nevertheless, we find by re-measuring a cutting's magnetic response after the room temperature XRD scan has finished, that some annealing occurs during the x-ray measurement. We do not believe this to have an influence on the results presented in this section, however as a precaution do not measure the same cutting more than once and each annealing time for SQUID and XRD measurements are from a different cutting of the pristine sample.

\begin{figure}
{	\includegraphics[width=0.9\linewidth]{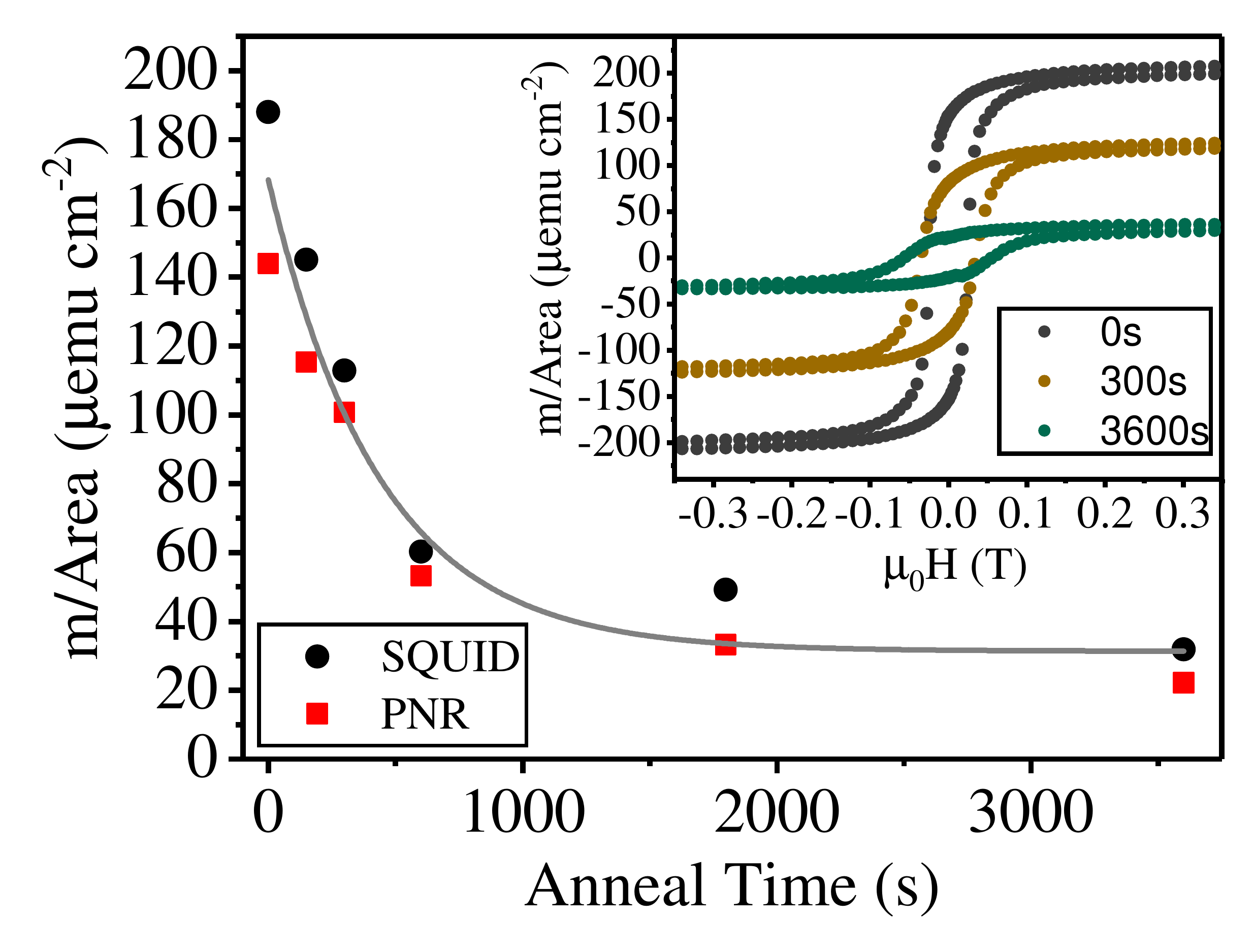}
}
  \caption{Magnetic characterization at 10~K of Si/SiO (100 nm)/Bi (50 nm)/Ni (6 nm)/Ta (5 nm) samples. The moment per area calculated from the hysteresis loops, inset, are shown with annealing time at 70\degC. The diamagnetic contribution due to the substrate has been subtracted. The uncertainty in moment/area is dominated by the area measurements (different cuttings of the sample are used for each annealing step), and is less than 5\% . Also shown is the moment/area extracted by fitting the PNR measurements. The line on the main figure is a guide for the eye.}
  \label{Fig:SQUID}
\end{figure}

The magnetic characterization of our Bi/Ni bilayer samples are shown in Figure \ref{Fig:SQUID}. We extract the saturation (applied field of 1~T) moment/area by measuring hysteresis loops for cuttings with different annealing times. Exemplar hysteresis loops are shown in the inset of Figure \ref{Fig:SQUID}. For the as-grown sample cutting, the saturation magnetization of the 4.4~nm Ni layer is 318\units{emu/cm^3}, reduced from the bulk value of 600\units{emu/cm^3}. The reduced magnetization is consistent with the formation of magnetic dead layers, which are often observed in thin film Ni \cite{Robinson2007}. For all sample cuttings, we find that a large magnetic field (nearly 1~T) is required to achieve full saturation, hence the hysteresis loops shown in Figure \ref{Fig:SQUID} inset do not fully close within the field range shown. As the film is annealed, two changes to the hysteresis loops are observed; firstly, the saturation moment/area of the sample reduces and secondly, the coercive field increases. Both observations are consistent with the PNR modeling which shows that as the sample is annealed non-magnetic \NiBi\ intermetallic forms reducing the ferromagnetic Ni thickness.

It is possible to compare the measured magnetic moment returned from the SQUID measurements and PNR fitting (Figure \ref{Fig:PNR}) by normalizing datasets to the areas of the samples. Collated moment/area with annealing time at 70\degC\ for the samples are shown Figure \ref{Fig:SQUID}. As the sample is annealed we observe an exponential type decay of the sample moment/area from about 188\units{emu/cm^{2}} in the as-grown state to 32\units{emu/cm^{2}} after 3600 s annealing at 70\degC. The time constant of the decay is 530\units{s}. The extracted moment/area by the two techniques show close agreement in both trend and absolute value for annealed samples, we note that there is some disagreement in the magnetic moment of the as-grown state.

\begin{figure}
    \includegraphics[width=0.9\linewidth]{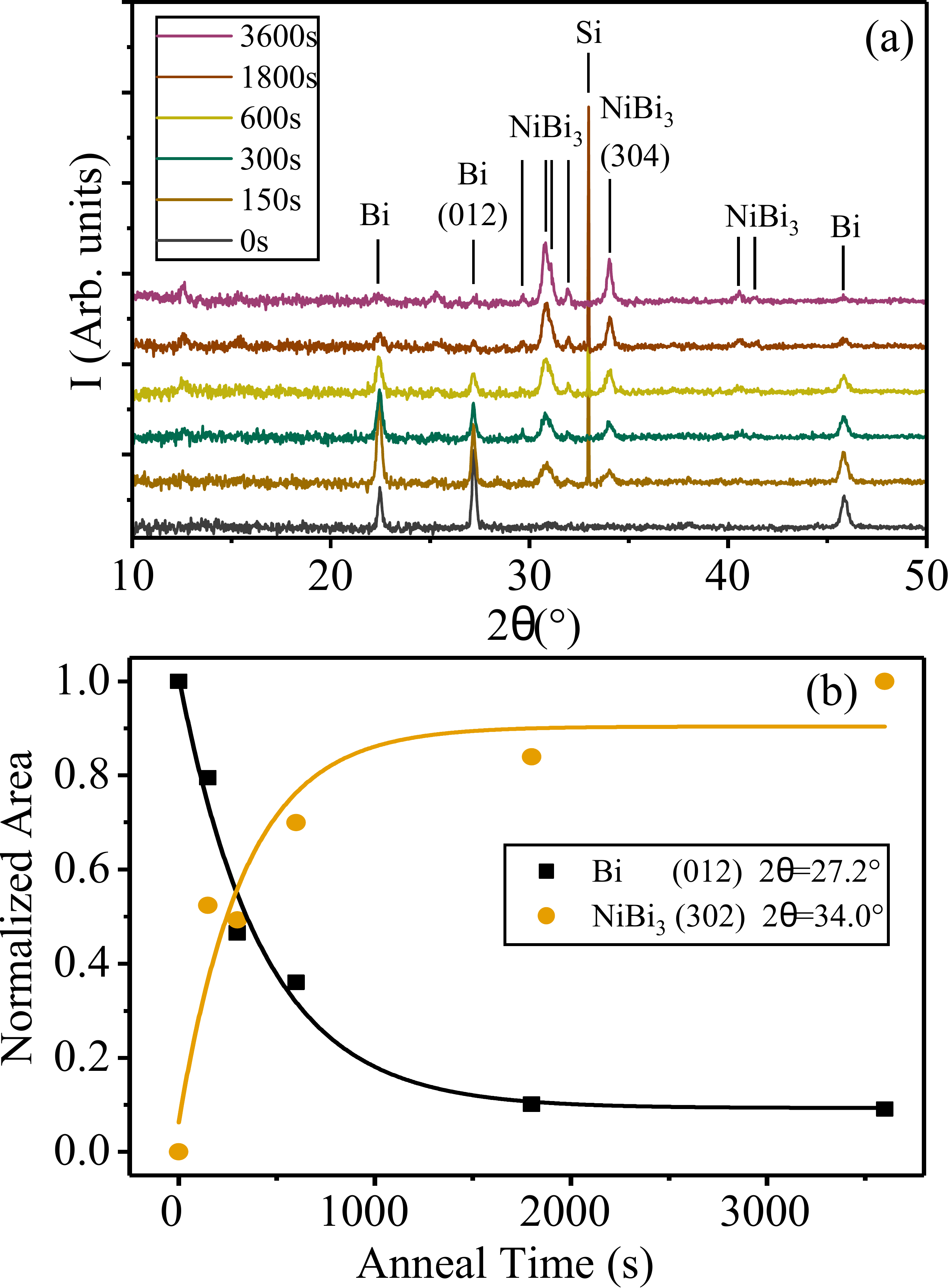}
  \caption{X-ray diffraction characterization of 70\degC~annealed Si/SiO (100 nm)/Bi (50 nm)/Ni (6 nm)/Ta (5 nm) samples. (a) Surfaced aligned XRD scans with annealing time. Peaks due to the Si substrate, Bi layer and \NiBi\ intermetallic are indexed. (b) The normalized area under the Bi (012) and \NiBi\ (302) peaks to show evolution of the sample with annealing step. Lines are a guide to the eye.}
  \label{Fig:XRD}
\end{figure}

Fig. \ref{Fig:XRD} (a) shows the results of XRD characterization. In the as-grown state, structural peaks due to the Bi layer and Si substrate are present in the sample. We do not expect Ni or Ta peaks to appear as these layers are too thin. After annealing, peaks appear in the XRD scan which correspond to the \NiBi\ intermetallic and equally the Bi peak intensity drops, suggesting that the textured Bi layer is being replaced by a textured \NiBi\ intermetallic. The timescale for these structural changes to occur to the sample is similar to the timescale where changes are observed in PNR and SQUID measurements annealed at the same temperature. The \NiBi\ has a weakly preferred orientations toward to (203).

\subsection{Superconducting Properties}

\begin{figure}
    \includegraphics[width=0.9\linewidth]{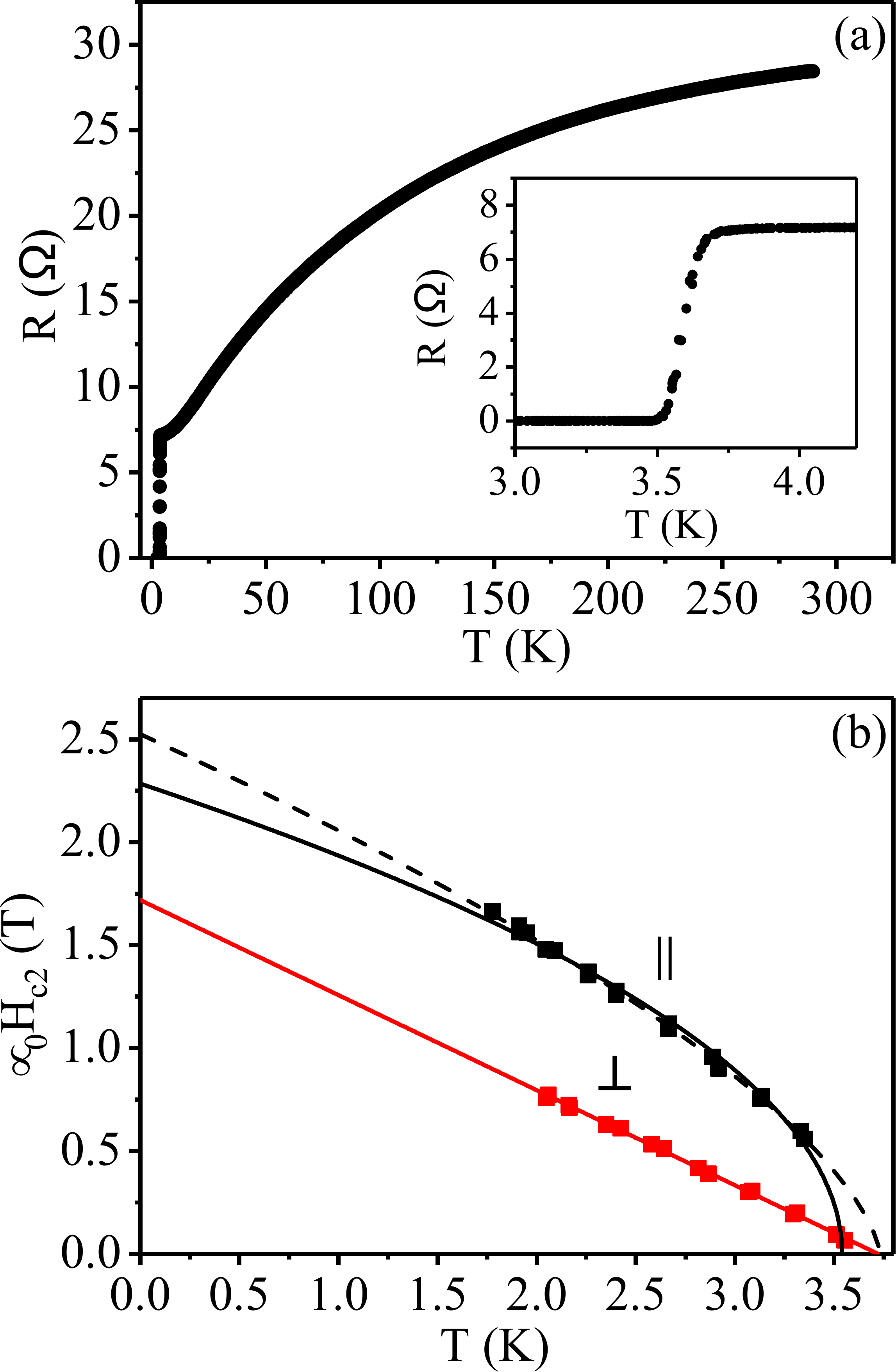}
  \caption{Electrical transport characteristic of Si/SiO (100 nm)/Bi (50 nm)/Ni (6 nm)/Ta (5 nm) sample after 70\degC~annealing for 3600 s. (a) Resistance verses temperature with the superconducting transition at 3.8\units{K} shown in the inset. (b) The out-of-plane (red) and in-plane (black) $H_{c2}$ data, solid lines a model fit for a thin superconductor (eq. \ref{2D_GL_OOP},\ref{2D_GL_IP}). The dashed line has the power of $\frac{1}{2}$ eq. \ref{2D_GL_IP} as a free fitting parameter and the \Tc\ is fixed to the \Tc\ measured by the RvT. $\mu _{0}H_{c2}^{\| }=2.28\units{T}$ and $\mu _{0}H_{c2}^{\bot }=1.72\units{T}$, the power of the dash line is 0.659.}
  \label{Fig:Critical}
\end{figure}

Low temperature 4-point probe transport measurements of a 70\degC\ fully annealed sample (3600 s) show a clear superconducting transition at 3.8\units{k}, Figure \ref{Fig:Critical} (a), the critical magnetic field in-plane and out-of-plane was also measured, Figure\ref{Fig:Critical} (b). The out-of-plane and in-plane $H_{c2}$ temperature dependence can be fitted effectively by the following Ginzburg-Landau (GL) model for a thin superconductor;

\begin{equation} \label{2D_GL_OOP}
\mu _{0}H_{c2}^\bot = \frac{\Phi _{\text{0}}}{2\pi \xi _{\text{GL}}(0)^2}(1-\frac{T}{T_{c}})
\end{equation}

\begin{equation} \label{2D_GL_IP}
\mu _{0}H_{c2}^\| =\frac{\Phi _{\text{0}} \sqrt{12}}{2\pi \xi _{\text{GL}}(0)d_{\text{SC}}} (1-\frac{T}{T_{c}})^{\frac{1}{2}}
\end{equation}

where $\Phi _{\text{0}}$, $\xi_{\text{GL}}(0) $ and $d_{\text{SC}}$ stands for flux quantum, in-plane coherence length and effective thickness of the superconductivity respectively. Taking the values from both in-plane and out-of-plane H$_{c2}$(0) fits the coherence length $\xi _{\text{GL}}$ as 13.8\units{nm} and the effective superconducting thickness $d_{\text{SC}}$ as 36.2\units{nm}, similar to the thickness of the \NiBi\ layer as obtained from the PNR data. The GL-theory for thin superconductors assumes that the $d_{\text{SC}}<\xi _{\text{GL}}$ which is not the case here and is visible in Figure \ref{Fig:Critical} (b) from the inadequate fitting for the in-plane $H_{c2}$ data.
Alternatively, making the power of the in-plane equation a free parameter (instead of a fixed $\frac{1}{2}$) returns a value of 0.659 and a more satisfactory fit to the experimental data. A returned power of 1 is expected for bulk behavior, suggesting this sample is in some intermediate state between bulk and thin superconductivity.

%
%
%
%
\subsection{Influence of Annealing Temperature}

\begin{figure}
    \includegraphics[width=0.9\linewidth]{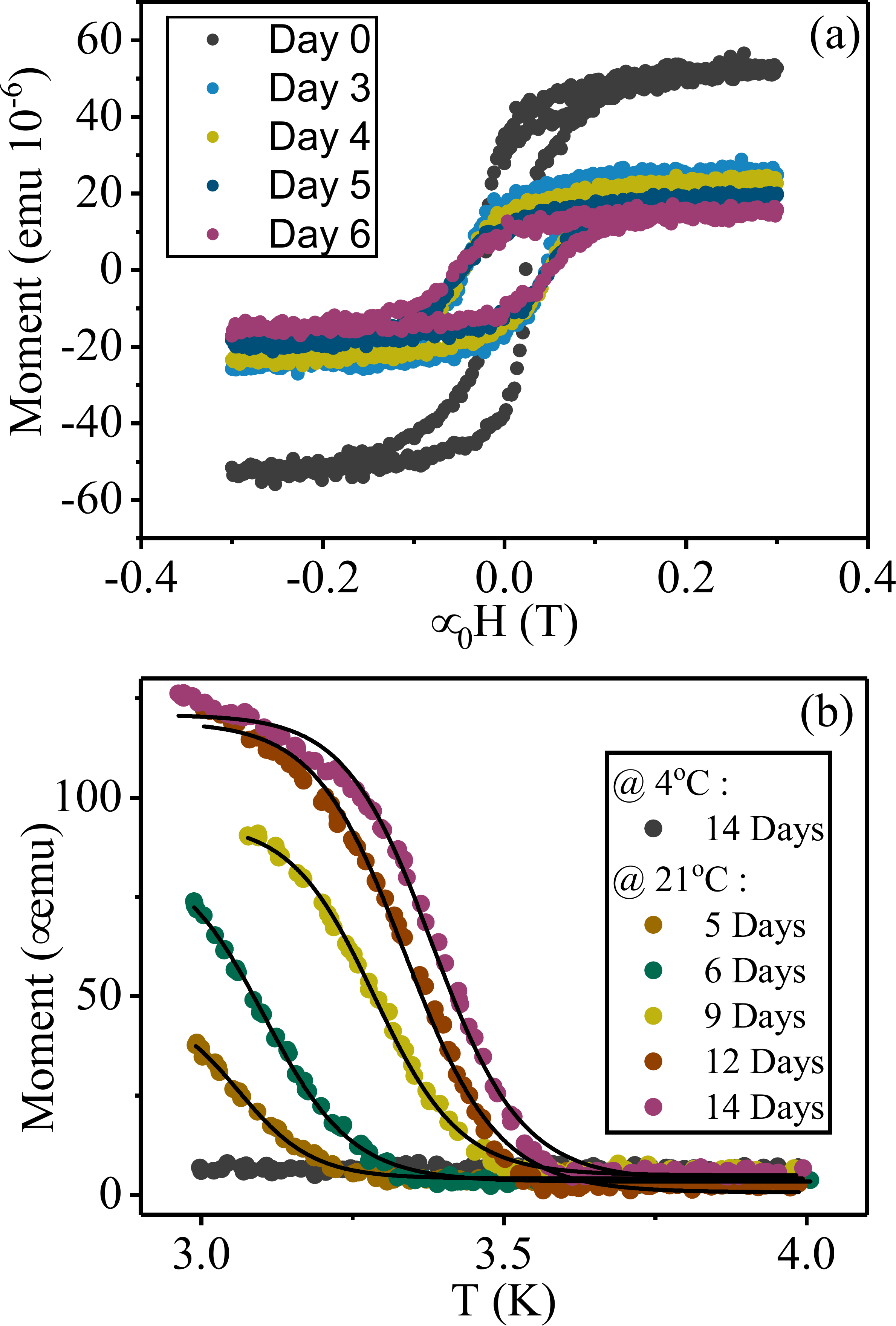}
  \caption{(a) Sample of Ta(5)/Ni(6)/Bi(50)\units{nm} stored at room temperature and hysteresis measured at 3\units{K}. (b) MvT of Ta(5)/Ni(4)/Bi(50)\units{nm} samples at two different temperatures, black lines are logistic function fitting for which the \Tc\ is taken at 20\% height.}
  \label{Fig:RoomTemp}
\end{figure}

Samples as-grown were not immediately observed to be superconducting, but after annealing or leaving at room temperature for several days the magnetization of the Ni layer was reduced although it is non vanishing and an increased coercivity (Fig.\ref{Fig:RoomTemp}.a), after that a superconducting transition appeared and increased until stabilizing at 3.8\units{k} (Fig.\ref{Fig:RoomTemp}.b). The thickness of the \NiBi\ layer grows quickly as the sample is annealed once the layer is thick enough to support a superconducting transition there is a meissner response that can be measured to deduce the \Tc . By measuring the \Tc\ at several points along the annealing process a time constant for the onset can be fitted from a decaying exponential. As long as the sample is far from saturating its \Tc\ the time constant will be not be affect by small amounts of annealing that may have already taken place. The inverse time constant against annealing temperature can be described well by the Arrhenius equation for a thermally activated reaction with an association activation energy of $(0.86\pm 0.06)\units{eV}$ (Fig.\ref{Fig:Anneal}.b). The activation energy is lower than for similar systems of interface mixing and diffusion with typical activation energies of $\approx$ 1\units{eV} \cite{nicolet1978,Hofler1993}.

\section{Discussion}\label{discussion}

\begin{figure}

    \includegraphics[width=0.9\linewidth]{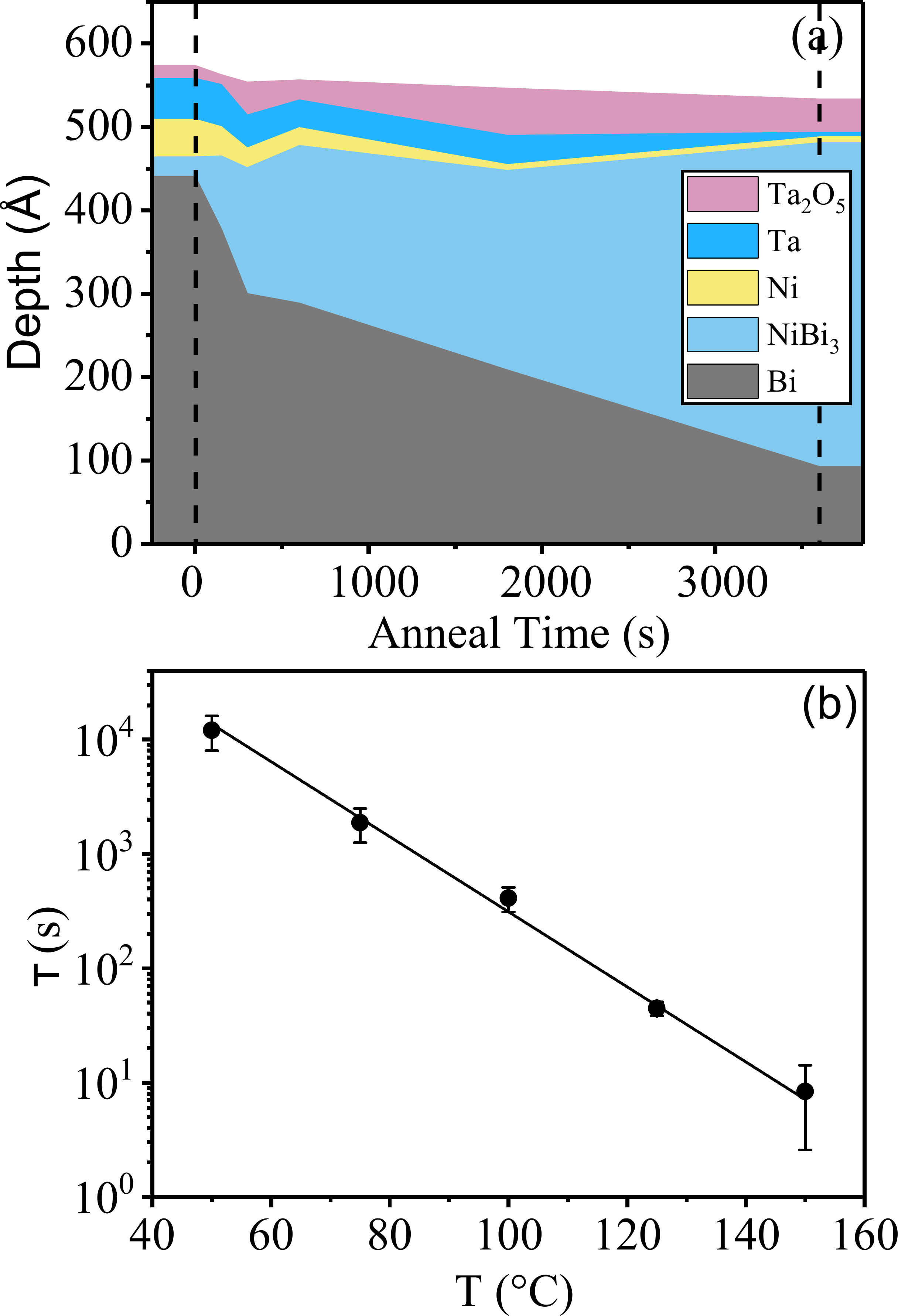}
    
  \caption{(a) Fitted thickness extracted from PNR data for each annealing step, omitting the roughness at the interface for simplicity. The annealing at 70\degC takes place between the dashed lines. (b) Arrhenius plot of Bi/Ni bilayers annealed in the temperature range of 50\degC -- 150\degC with an activation energy of $(0.86\pm 0.06)\units{eV}$, time constants taken from the superconducting \Tc\ onset rate.}
  \label{Fig:Anneal}
\end{figure}

Figure.\ref{Fig:Anneal} (a) shows the fitted thickness for each annealing step as extracted from fitting to PNR. The Ta$_{2}$O$_{5}$ increases as expected for heating in air and has protected the lower layers from oxidation. The thickness of the Bi and Ni layers reduce as the \NiBi\ layer becomes thicker. We also observe an overall reduction in film thickness as the \NiBi\ alloy is denser than the individual Bi or Ni layers. To optimize the \Tc\ 50\units{nm} of Bi and 6\units{nm} of Ni are used, the ratio maybe suggest that to obtain the highest \Tc\ annealed samples one should minimize the remaining Ni and maximize the thickness of the \NiBi \cite{NoteSM}.

Extrapolating from the Arrhenius plot at 21\degC\ the time constant for the annealing rate will be 4\units{days} which is comparable to the timescale measured at room temperature, the exponential has a doubling rate of $\approx$ 6\degC\ such that if kept at 3\degC\ the timescale extends to 12\units{days} (Fig.\ref{Fig:Anneal}). The low annealing temperatures can be understood from the low melting point of Bi of 544\units{K} being about twice room temperature ($\approx 294\units{K}$). It is not uncommon to anneal thin films to form alloys at 50\% of the constituent layers melting points . The Bi atoms become mobile at higher temperatures diffusing across the interface and the same for Ni as it diffuses into the Bi layer forming a \NiBi\ alloy.

When first measured in PNR, the as-grown sample has a thin layer at the interface with a density similar to \NiBi\ , although in the corresponding XRD the \NiBi\ peaks are at the limit of signal to noise and so the volume fraction of \textit{ordered} \NiBi\ is very small. In comparison, in samples that have been intentionally annealed, the ratio for \NiBi\ XRD peak areas and the fitted layer thickness suggest that 2\units{nm} of \NiBi\ should have a greater intensity of XRD peaks than is measured in the as-grown sample. This thin intermixed layer is not unexpected. The roughness of the Bi layer, implantation of Ni adatoms during growth and, probably most significantly, the time taken to get the samples from a room temperature vacuum chamber to the freezer all contribute to some intermixing of the interface. 

Liu \textit{et al} perform pulsed laser deposition (PLD) of Bi/Ni bilayers where they report the \NiBi\ alloy forms during growth. Their interpretation is that PLD is a nonequilibrium process such that Ni atoms arrive at the Bi layer with enough energy to implant deep into the film. Ni implantation stops at the SiO$_{2}$, where it accumulates forming a NiBi alloy and a \NiBi\ layer further from the SiO$_{2}$. We find that ordered \NiBi\ does not exist in our samples at growth, but forms during annealing and is confined to the Bi/Ni interface. We do not find evidence for ordered NiBi alloy in our films, however in the final annealed state Bi may have contaminated any remaining Ni layer in the structure. We conclude that during growth the Ni implantation depth is confined to near the top surface of the Bi, as fitted by the 2\units{nm} of \NiBi\ in the as-grown state.

By measuring PNR below \Tc\ in an applied field, it is possible to observe Meissner screening under the right conditions (that the film thickness, roughness, and the superconducting penetration depth are balanced to allow for measurable screening to occur) \cite{PhysRevB.52.10395,PhysRevB.80.134510}. It is also possible that exotic superconducting states can influence the magnetic response of a superconducting sample \cite{Luke97,flokstra2016remotely,flokstra2019manifestation}. Here, we measure both the as-grown and fully annealed states of the sample at 3~K (below the superconducting transition for the annealed sample) to look for changes to magnetic response from the sample. No such changes are observed below \Tc\ in the PNR, most likely as the \NiBi\ superconductor is too thin to observe Meissner screening.

Experimental results by other techniques suggest that superconductivity in bilayers of Bi/Ni may be spin-triplet ($p$-wave) in nature \cite{Gong15,Gong16,Wang17,Zhou16,Zhou17,Chauhan19,tokuda2019spin}. The symmetric spin-triplet states are found in only a handful of superconductors, where Sr$_2$RuO$_4$ is currently the best candidate \cite{Luke97}. In these materials, the antisymmetry requirements are satisfied by the condensates of these superconducting materials being spatially antisymmetric, that is, odd in angular momentum. It is also possible to generate spin-triplet states in proximity coupled thin films where $s$-wave pairing is retained by introducing a Berizinskii state with spontaneous breaking of time-reversal symmetry \cite{PhysRevLett.86.4096}. Our Bi/Ni samples contain all the necessary ingredients for such a state to occur; a source of $s$-wave superconductivity (\NiBi ), ferromagnetism (Ni) and strong spin-orbit coupling (Bi) \cite{PhysRevB.89.134517,jeon2018enhanced,satchellSOC2018,PhysRevB.99.174519}.

\section{Conclusions}

In this work samples of Bi/Ni bilayers are observed initially to be non-superconducting until either left at room temperature for several days or a short period of heating as low as $+50\degC$. The superconductivity is attributed to intermixing of the Bi/Ni interface forming an alloy of \NiBi\ identified by XRD Bragg peaks. PNR data is consistent with initially distinct Bi/Ni layers with minimal \NiBi\ in the as-grown states that when annealed at $70\degC$ for 1\units{hour} the diffusion across the interface increases the \NiBi\ thickness and reduces the pure Bi/Ni layers.

From this it seems that to properly study clean and distinct interfaces that maintaining a low temperature for preparation and storage is important. Normal device fabrication recipes which use heating to bake resist or growth methods that allow the sample to heat up will fully anneal Bi/Ni bilayers. Although the superconductivity origins from a known bulk superconductor the existence is still interesting as it is in close proximity to a ferromagnetic layer, a strong spin-orbit coupling Bi layer and also the likely proximity effect in the Bi layer. \\


The data associated with this paper are openly available from the University of Leeds and ISIS Neutron and Muon source data repositories \cite{NoteY}.

\begin{acknowledgments}
We thank the ISIS neutron and muon source for allocating beamtime (RB:1920455). The work was supported financially through the EPSRC Doctoral Training Partnership and grant EP/M000923/1.  This project has received funding from the European Unions Horizon 2020 research and innovation programme under the Marie Sk\l{}odowska-Curie Grant Agreement No. 743791 (SUPERSPIN).

\end{acknowledgments}

\bibliography{BiNi_Ref_3}

\end{document}


\begin{center}
\large{Supplementary Material for: ``On the Origin of Superconductivity at Nickel-Bismuth Interfaces'' by Matthew Vaughan \textit{et al.}}
\end{center}


\section{Sputter Deposition Details}

Samples were grown by DC sputtering from pure metal targets of bismuth (99.99\%) and nickel (99.95\%) at $4.3$\units{\AA s^{-1}} and $3$\units{\AA s^{-1}} respectively. The substrate is thermally oxidized Si with 100\units{nm} of SiO$_{2}$ cleaned using 5\units{mins} of acetone and then isopropyl ultrasonic cleaning. Using a pure argon (6 nines) atmosphere the growth pressure is $ 0.43\units{Pa}$ for Bi and $ 0.61\units{Pa}$ for Ni, the pressure distance product is $ 4.4\units{Pa~cm}$ and $ 3.8\units{Pa~cm}$ respectively. Each target is pre-sputtered for 5\units{mins} as well as a Nb target to act as a getter, a liquid nitrogen meissner trap also reduces the partial pressure of water. The typical base pressure is $ 1.2\times 10^{-5}\units{Pa}$ and temperature is $21\degC$. Growth rates are calibrated by fitting to Keissig fringes obtained by low-angle x-ray reflectometry on single layer reference samples.

After growth, samples were stored in a domestic freezer (Beko fridge/freezer model CDA543FW) at $-20\units{^\circ C}$ and transported in a domestic portable refrigerator (Halfords 24l 12V Electric Cooolbox) at $\approx 4 \units{^\circ C}$. Samples were vacuum packing in airtight plastic to minimize condensation. 31 days elapsed between sample growth and the PNR measurements.

\section{Optimization of the Superconducting Phase}

We find that to maximize the \Tc\ an optimal of 50\units{nm} of bismuth and 6\units{nm} of nickel is used.

\begin{figure}[H]
  \includegraphics[width=0.9\linewidth]{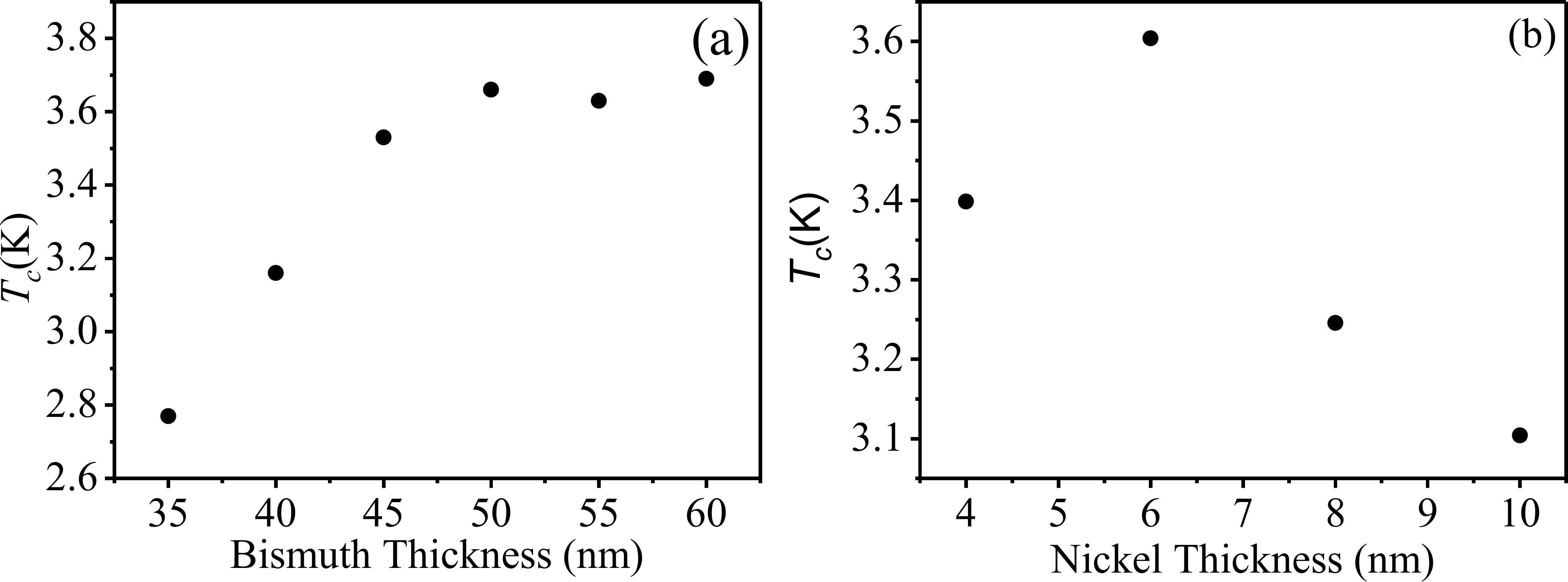}
  \caption{Optimization of superconducting \Tc\ . (a) Nickel layer fixed at 6\units{nm} with bismuth layer varied, after 2 weeks at room temperature.(b) Bismuth layer fixed at 50\units{nm} and nickel layer varied, annealed at $100\degC$. Maximize above 50\units{nm} of bismuth and 6\units{nm} of nickel.}
  \label{Fig:Thickness}
\end{figure}

\pagebreak

\section{PNR Box Model Fitting values}

The full set of GenX fitting parameters for a Ta$_{2}$O$_{5}$/Ta/Ni/\NiBi\ /Bi/SiO$_{2}$/Si box model. The Final two annealing steps can be fitted better by a NiBi layer in place of a pure Ni layer. The figure of merit (FOM) for each annealing step is shown, with a NiBi layer on the last two the FOM decrease to 1.167 and 1.807 for 1800\units{s} and 3600\units{s} respectively.

\begin{table}
  \centering
  \begin{tabular}{|l|l|l|l|l|l|l|}
	\hline
	Parameter & 0\units{s} & 150\units{s} & 300\units{s} & 600\units{s} & 1800\units{s} & 3600\units{s} \\ \hline
	
	FOM & 1.421 & 1.341 & 1.415 & 1.404 & 1.174 & 1.877 \\ \hline \hline

	Ta$_{2}$O$_{5}$: \textit{d} (nm) & 1.6 $\pm $ 0.3 & 1.2 $\pm $ 0.2 & 3.9 $\pm $ 0.8 & 2.4 $\pm $ 0.5 & 5.7 $\pm $ 0.8 & 4.0 $\pm $ 0.5 \\ \hline
	$\rho $ (\% bulk) & 95 $\pm $ 6 & 102 $\pm $ 5 & 85 $\pm $ 9 & 89 $\pm $ 5 & 83 $\pm $ 3 & 118 $\pm $ 8 \\ \hline
	$\sigma $(nm) & 3.8 $\pm $ 0.2 & 4.0 $\pm $ 0.1 & 4.4 $\pm $ 0.3 & 4.2 $\pm $ 0.3 & 4.4 $\pm $ 0.4 & 5.0 $\pm $ 0.2 \\ \hline \hline

	Ta: \textit{d} (nm) & 4.9 $\pm $ 0.4 & 5.1 $\pm $ 0.7 & 3.9 $\pm $ 0.6 & 3.3 $\pm $ 0.3 & 3.5 $\pm $ 0.5 & 0.5 $\pm $ 0.5 \\ \hline
	$\rho $ (\% bulk) & 92 $\pm $ 1 & 104 $\pm $ 3 & 103 $\pm $ 6 & 97 $\pm $ 5 & 103 $\pm $ 5 & 80 $\pm $ 10 \\ \hline
	$\sigma $(nm) & 18 $\pm $ 0.3 & 4.2 $\pm $ 0.6 & 3.8 $\pm $ 0.1 & 1.5 $\pm $ 0.9 & 4.1 $\pm $ 0.3 & 2.3 $\pm $ 0.4 \\ \hline \hline

	Ni: \textit{d} (nm) & 4.5 $\pm $ 0.7 & 3.5 $\pm $ 0.1 & 2.4 $\pm $ 0.1 & 2.1 $\pm $ 0.1 & 0.7 $\pm $ 0.2 $\dagger$ & 0.7 $\pm $ 0.1 $\dagger$ \\ \hline
	$\rho $ (\% bulk) & 91 $\pm $ 1 & 83 $\pm $ 2 & 90 $\pm $ 1 & 101 $\pm $ 3 & 90 $\pm $ 20 & 87 $\pm $ 8 \\ \hline
	$\sigma $(nm) & 3.2 $\pm $ 0.2 & 2.7 $\pm $ 0.2 & 3.9 $\pm $ 0.2 & 3.3 $\pm $ 0.2 & 3.9 $\pm $ 0.2 & 4.0 $\pm $ 0.2 \\ \hline
	M ($\mu_B$/atom) & 0.40 $\pm $ 0.02 & 0.46 $\pm $ 0.03 & 0.51 $\pm $ 0.03 & 0.45 $\pm $ 0.03 & 0.57 $\pm $ 0.04 & 0.27 $\pm $ 0.09 \\ \hline \hline

	\NiBi\ : \textit{d} (nm) & 2.3 $\pm $ 0.4 & 8.7 $\pm $ 0.7 & 15 $\pm $ 2 & 19 $\pm $ 2 & 24 $\pm $ 1 & 38.8 $\pm $ 0.7 \\ \hline
	$\rho $ (\% bulk) & 92 $\pm $ 2 & 103 $\pm $ 3 & 99 $\pm $ 1 & 96 $\pm $ 7 & 99 $\pm $ 2 & 91 $\pm $ 1 \\ \hline
	$\sigma $(nm) & 4.7 $\pm $ 0.3 & 4.5 $\pm $ 0.2 & 3.5 $\pm $ 0.3 & 4.6 $\pm $ 0.3 & 3.8 $\pm $ 0.3 & 5.2 $\pm $ 0.3 \\ \hline
	M ($\mu_B$/atom) & 0.03 $\pm $ 0.1 & 0.01 $\pm $ 0.05 & 0.01 $\pm $ 0.02 & 0.01 $\pm $ 0.02 & 0 $\pm $ 0.03 & 0 $\pm $ 0.01 \\ \hline \hline

	Bi: \textit{d} (nm) & 44.1 $\pm $ 0.4 & 37.8 $\pm $ 0.8 & 30 $\pm $ 1 & 29 $\pm $ 2 & 20.9 $\pm $ 0.8 & 9.3 $\pm $ 0.7 \\ \hline
	$\rho $ (\% bulk) & 96 $\pm $ 1 & 95 $\pm $ 2 & 97 $\pm $ 2 & 97 $\pm $ 1 & 106 $\pm $ 3 * & 105 $\pm $ 1 * \\ \hline
	$\sigma $(nm) & 7.9 $\pm $ 0.9 & 9.5 $\pm $ 0.7 & 7.9 $\pm $ 0.8 & 7 $\pm $ 2 & 15 $\pm $ 3 & 9 $\pm $ 1 \\ \hline \hline

	SiO$_{2}$: \textit{d} (nm) & 95.0 $\pm $ 0.3 & 95.1 $\pm $ 0.6 & 95.1 $\pm $ 0.3 & 95.1 $\pm $ 0.3 & 94.7 $\pm $ 0.4 & 93.9 $\pm $ 0.5 \\ \hline
	$\rho $ (\% bulk) & 100.5 $\pm $ 0.1 & 99.8 $\pm $ 0.2 & 99.9 $\pm $ 0.7 & 99.9 $\pm $ 0.3 & 100.6 $\pm $ 0.3 & 100.1 $\pm $ 0.2 \\ \hline
	$\sigma $(nm) & 0.1 $\pm $ 0.4 & 0.1 $\pm $ 0.7 & 0.2 $\pm $ 0.5 & 0.4 $\pm $ 0.6 & 0.2 $\pm $ 0.4 & 0.2 $\pm $ 0.6 \\ \hline \hline

	Si: $\rho $ Fixed (\% bulk) & 100 & 100 & 100 & 100 & 100 & 100 \\ \hline
	$\sigma $(nm) & 0.1 $\pm $ 0.3 & 0.1 $\pm $ 0.4 & 0.2 $\pm $ 0.3 & 0.2 $\pm $ 0.3 & 0.2 $\pm $ 0.2 & 0.3 $\pm $ 0.2 \\ \hline
	\end{tabular}

  \caption{Fitting parameters for the each $70\degC$~annealing step with the FOM. The Si substrate density is fixed to the bulk value and all other parameters are free fitting. The density for each free layer is fitted with tight bounds to remain physical. (*) The final two Bi densities are higher that bulk as there is Ni inclusion throughout the layer. ($\dagger$ ) GenX does not correctly fit very thin layers}
  \label{Fig:GenX}
\end{table}

\begin{table}
\begin{tabular}{|l|l|l|}
\hline
Parameters & 1800s & 3600s \\ \hline

FOM & 1.167 & 1.807  \\ \hline \hline

Ta$_{2}$O$_{5}$: \textit{d} (nm) & $2.2\pm 0.4$ & $5.1\pm 0.8$ \\ \hline
$\rho $ (\% bulk) & $90\pm 10$ & $79\pm 2$ \\ \hline
$\sigma $(nm) & $4.4\pm 0.2$ & $4.2\pm 0.1$ \\ \hline \hline

Ta: \textit{d} (nm) & $2.2\pm 0.4$ & $3.1\pm 0.5$ \\ \hline
$\rho $ (\% bulk) & $101\pm 7$ & $103\pm 2$ \\ \hline
$\sigma $(nm) & $4\pm 2$ & $1\pm 2$ \\ \hline \hline

NiBi: \textit{d} (nm) & $3.6\pm 0.8$ & $3.5\pm 0.8$ \\ \hline
$\rho $ (\% bulk) & $108\pm 9$ & $90\pm 3$ \\ \hline
$\sigma $(nm) & $4\pm 2$ & $10\pm 1$ \\ \hline
M ($\mu_B$/atom) & $0.4\pm 0.1$ & $0.5\pm 0.2$ \\ \hline \hline

\NiBi\ : \textit{d} (nm) & $24\pm 1$ & $38\pm 1$ \\ \hline
$\rho $ (\% bulk) & $98\pm 2$ & $88.4\pm 0.8$ \\ \hline
$\sigma $(nm) & $5.5\pm 0.6$ & $8.5\pm 0.6$ \\ \hline
M ($\mu_B$/atom) & $0.02\pm 0.02$ & $0.00\pm 0.03$ \\ \hline \hline

Bi: \textit{d} (nm) & $22\pm 1$ & $4.5\pm 0.4$ \\ \hline
$\rho $ (\% bulk) & $106\pm 2$ & $92\pm 2$ \\ \hline
$\sigma $(nm) & $16\pm 1$ & $3.6\pm 0.5$ \\ \hline \hline

SiO$_{2}$: \textit{d} (nm) & $94.4\pm 0.3$ & $94.9\pm 0.2$ \\ \hline
$\rho $ (\% bulk) & $100.7\pm 0.2$ & $100.0\pm 0.3$ \\ \hline
$\sigma $(nm) & $0.1\pm 0.5$ & $0.1\pm 0.4$ \\ \hline \hline

Si: $\rho $ Fixed (\% bulk) & 100.0 & 100.0 \\ \hline
$\sigma $(nm) & $0.1\pm 0.3$ & $0.1\pm 0.3$ \\ \hline

\end{tabular}
  \caption{With the NiBi layer instead of the Ni layer, bulk density from the ordered alloy of NiBi.}
  \label{Fig:NiBi}
\end{table}